# Mirror production for the Cherenkov telescopes of the ASTRI Mini-Array and of the MST project for the Cherenkov Telescope Array

N. La Palombara[1,*], G. Sironi[2], E. Giro[3], S. Scuderi[1], R. Canestrari[4], S. Iovenitti[2,5], M. Garczarczyk[6], M. Krause[6], S. Diebold[7], R. Millul[2], F. Marioni[8], N. Missaglia[8], M. Redaelli[8], G. Valsecchi[8], F. Zocchi[8], A. Zanoni[9], and G. Pareschi[2]
*for the ASTRI[10] and the CTA[11] projects*
[1]INAF - IASF Milano, Via A. Corti 12, Milano, Italy, I-20133
[2]INAF - Osservatorio Astronomico di Brera, Via Bianchi 46, Merate, Italy, I-23807
[3]INAF – Osservatorio Astronomico di Padova, Vicolo Osservatorio 5, Padova, Italy, I-35122
[4]INAF - IASF Palermo, Via U. La Malfa 153, Palermo, Italy, I-90146
[5]Dipartimento di Fisica, Università degli Studi di Milano, via Celoria 16, Milano, Italy, I-20133
[6]Deutsches Elektronen-Synchrotron (DESY), Platanenallee 6, Zeuthen, Germany, D-15738
[7]Institut für Astronomie und Astrophysik Tübingen (IAAT), Sand 1, Tübingen, Germany, D-72076
[8]Media Lario s.r.l., Via al Pascolo, Bosisio Parini, Italy, I-23842
[9]ZAOT s.r.l., Via E. Restelli 20, Vittuone, Italy, I-20010
[10]http://www.astri.inaf.it/
[11]https://www.cta-observatory.org/

**Abstract**. The Cherenkov Telescope Array (CTA) is the next ground-based gamma-ray observatory in the TeV γ-ray spectral region operating with the Imaging Atmospheric Cherenkov Technique. It is based on almost 70 telescopes of different class diameters – LST, MST and SST of 23, 12, and 4 m, respectively - to be installed in two sites in the two hemispheres (at La Palma, Canary Islands, and near Paranal, Chile). Several thousands of reflecting mirror tiles larger than 1 m$^2$ will be produced for realizing the segmented primary mirrors of a so large number of telescopes. Almost in parallel, the ASTRI Mini-Array (MA) is being implemented in Tenerife (Canary Islands), composed of nine 4 m diameter dual-mirror Cherenkov telescopes (very similar to the SSTs). We completed the mirror production for all nine telescopes of the ASTRI MA and two MST telescopes (400 segments in total) using the cold glass slumping replication technology. The results related to the quality achieved with a so large-scale production are presented, also discussing the adopted testing methods and approaches. They will be very useful for the adoption and optimization of the quality assurance process for the huge production (almost 3000 m$^2$ of reflecting surface) of the MST and SST CTA telescopes.

**\*** E-mail: nicola.lapalombara@inaf.it

## 1 Introduction

Imaging Atmospheric Cherenkov Telescopes (IACTs) are ground-based telescopes designed to observe γ-ray celestial sources at very-high energies (VHE), in the TeV energy range [1]. After the first detection of a VHE source, obtained in 1989 with the *Whipple* telescope [2], several other sources have been found over the last three decades, thanks to the *H.E.S.S.* [3], *MAGIC* [4], and *VERITAS* [5] telescopes. Currently, more than 200 VHE sources are known[*]: some of them have been identified with different classes of celestial objects, but a significant fraction of VHE sources are still unidentified [6]; in addition, due to the limited sensitivity of the current generation of instruments, in most cases the origin of the observed VHE emission is not fully understood. Therefore, new facilities with better performance are necessary.

---

[*] http://tevcat.uchicago.edu/

ASTRI is a project led by the Italian National Institute for Astrophysics (INAF) that aims to realize a series of dual-mirror IACTs. They are characterized by innovative technological solutions, such as the Schwarzschild-Couder optical configuration [7], a modular, light and compact focal-plane camera consisting of an array of multi-pixel silicon photo-multiplier sensors, and an efficient and fast front-end electronics, specifically designed for ASTRI. The feasibility of this innovative design has been successfully demonstrated with the *ASTRI-Horn* prototype. This telescope, installed in 2014 at the Serra La Nave site on Mount Etna (Catania, Italy), is the first Schwarzschild-Couder telescope to be built and tested [8]: it observed its first light in May 2017[†] and detected the Crab nebula in December 2018 [9].

On behalf of INAF, the ASTRI Collaboration is developing a Mini-Array (ASTRI MA) composed of nine dual-mirror IACTs, based on the design of the *ASTRI-Horn* prototype [10, 11]. They will be installed at the Teide Astronomical Observatory, operated by the Instituto de Astrofisica de Canarias (IAC), on Mount Teide, in the Canary island of Tenerife. The ASTRI MA will be operated by INAF based on a host agreement with IAC[‡] [12].

In addition, the ASTRI Collaboration is participating to the preparatory effort for the realization of the Cherenkov Telescope Array (CTA). It will be located in two different sites, one in the northern and one in the southern hemisphere, and will be made up of three different types of IACTs [13]:
- the Large-Sized Telescopes (LSTs), with a 23 m mirror diameter, will cover the low-energy range between 20 GeV and 3 TeV;
- the Medium-Sized Telescopes (MSTs), with a 12 m mirror diameter, will observe the core energy range between 80 GeV and 50 TeV;
- the Small-Sized Telescopes (SSTs), with a 4 m primary mirror diameter, will be devoted to the high-energy range from ~ 1 TeV up to more than 300 TeV.

With respect to the initial design configuration that was based on almost 100 telescopes [14], the CTAO arrays have recently been descoped: the baseline configuration, dubbed as "Alpha Configuration", will consist of a Northern Array with 4 LSTs and 9 MSTs and a Southern Array with 14 MSTs and 37 SSTs.

The improvement in the on-axis differential sensitivity of the two CTAO Arrays with respect to the currently existing instruments of the same kind (mainly MAGIC, H.E.S.S., and VERITAS) is a factor 5 to 10 depending on the energy range[§] [15]. Particularly relevant is the CTAO improvement in terms of off-axis sensitivity. Between 3 and 4 degrees away from the pointing direction the differential sensitivity worsens just by less than a factor 2 at all energies with respect to the on-axis sensitivities. CTAO has therefore a γ-ray Field-of-View (FoV) of seven degrees in diameter at 1 TeV (even more at higher energies).

It should be noted that the performance of CTAO would also be improved, after the implementation of this "Alpha Configuration", with the installation of additional telescopes, in particular for the southern site, where there is a large area that could be populated with additional telescopes. In this respect, it should be noted that another kind of medium size telescopes, based on the dual mirror Schwarzschild-Couder aplanatic telescopes (9 m in diameter) and named pSCT, would be possibly implemented, similar to the prototype already developed and operated in Arizona (USA) [16, 17].

---

[†] https://www.cta-observatory.org/cta-prototype-telescope-astri-achieves-first-light/
[‡] http://www.inaf.it/en/inaf-news/astri-a-new-pathfinder-of-the-arrays-of-cherenkov-telescopes
[§] Public CTAO webpage reporting a full set of performance plots for the Alpha Configuration can be found at https://www.cta-observatory.org/science/ctao-performance/

In the context of CTA, INAF is leading the international collaboration [18, 19] that will provide, in terms of in-kind contribution to CTAO, the SST telescopes. The design will be based on that of the *ASTRI-Horn* prototype and of the ASTRI MA telescopes. The mirrors will be provided by INAF and they will be of the same type of the ASTRI MA.

The MSTs are based on a single-mirror Davies-Cotton optical design [20]. Each of the 12 m primary mirrors will be made assembling together 86 identical hexagonal segments with spherical profile and a maximum size of 120 cm. INAF, in the context of the ASTRI collaboration, is also responsible for the procurement of all the reflecting segments for the 9 MSTs that will be implemented at the Northern site (800 segments to be produced, including the spares). The sandwiched mirror segments are produced using the so-called glass cold-slumping replication technology, initially set-up by INAF - Osservatorio Astronomico di Brera (OAB) and Media Lario (ML) company (Bosisio Parini, Italy) for the MAGIC II telescope [21] and then specifically developed to fit the requirements and specifications of the MSTs [22]. In this respect, ASTRI is cooperating with the Deutsches Elektronen-Synchrotron (DESY, Germany), responsible for the MST structures.

Also, the reflecting segments of the primary mirrors for the ASTRI MA are produced using the same technology [23] and they will be used for the production of segments for the primary mirrors of the 37 (extendable to 40) SST telescopes to be installed at the southern site (with about other 800 segments to be produced, including the spares). Each primary is based on 18 hexagonally-shaped segments, with a maximum size of 85 cm.

At the end of 2019, we completed the mirror production for both the nine telescopes of the ASTRI MA (200 reflecting segments of three different curvature radii and profile, corresponding to the three coronas of segments to form a primary) and two MST telescopes (with 200 identical reflecting segments with spherical shape produced). In this effort the ML Company represents the industrial partner, while ZAOT (Vittuone, Italy) is involved for the application of the reflective coating on the substrates. In this context, we have performed independent verifications of the optical performance of representative mirror samples.

In this paper, the aspects related to the large-scale production of these segments are presented, with a particular regard to the qualification activities that have been performed in order to assess and consolidate the production process. In this respect, the criteria adopted for the quality assurance, in order to monitor and verify the production reliability, are discussed in terms of performance and compliance with respect to the input requirements. Different methods and techniques have been used to perform the verifications. Finally, we present the performance of the produced segments and discuss their compliance and production yield with respect to the input requirements. The results of the present work will be very useful for the adoption and optimization of the quality assurance process for the huge production of the MST north and SST CTA telescopes, for which thousands of reflecting segments have to be manufactured and qualified.

## 2   Optical Design

*2.1 The Optical Design of the telescopes for ASTRI MA*

The dual-mirror telescopes of the ASTRI MA are based on the polynomial Schwarzschild-Couder optical design, an aplanatic layout that, thanks to the combination of the two mirrors, is characterized by a wide field of view, a low vignetting and isochrony. The plate scale of the ASTRI telescopes is 37.6 mm/°.

The primary mirror (ASTRI M1) has a radial symmetry and its optical profile is described by an aspherical polynomial function (of order 10), where the radius of the spherical component is 8.223 m. It has a diameter of 4.3 m and is segmented into 18 hexagonal segments; the face-to-face dimension of each segment is 85 cm. They are distributed in three concentric coronae (COR1, COR2, and COR3) of six identical segments each, where each corona has a different radial distance from the telescope axis. In fact, the best-fit radius of curvature (RoC) of the segments is 8.6, 9.8, and 11.7 m, respectively for the inner, medium, and outer corona [24].

The secondary mirror (ASTRI M2) is a monolithic element and has a diameter of 1.8 m. It has an aspherical shape, with a radius of 2.18 m for the spherical component, and is installed at a distance of 3.108 m from ASTRI M1. This implies an equivalent focal length for the whole telescope of 2.15 m [18].

*2.2 The Optical Design of the Single-Mirror MST*

The medium-sized telescope has a single-mirror configuration with a modified Davies-Cotton optical design [25], where the optical parameters are optimized to reduce the time spread at the Cherenkov camera focal plane from light reflected at different mirror positions. The mirror is segmented into 86 hexagonal segments, for a diameter of 12 m. All segments are identical: they have a spherical profile, with a RoC of 32.14 m, and the face-to-face size is 1.2 m. The telescope focal length is 16 m and the mirror segments are aligned to reflect rays parallel to the optical axis into the focal point. They are mounted on a spherical dish with a radius of 19.2 m. The plate scale of the MST telescopes is 280 mm/°.

## 3  Mirror Design and Production

In the framework of CTA several different methods have been considered for the mirror realization, which are based on different technologies [26, 27, 28, 29]: they include sandwich concepts with cold-slumped surfaces made of thin float glass and different core materials (like aluminum honeycomb, glass foams or aluminum foams), constructions based on carbon fiber/epoxy or glass fiber substrates, as well as sandwich structures made entirely of aluminum. In addition, lightweight polished aluminum mirrors with Ni/P alloy ("electroless Nickel") cladding as the reflecting surface were proposed for the Gamma-ray Cherenkov Telescope (GCT), the other dual-mirror telescope with a Schwarzschild-Couder optical design [30] that was proposed for the SSTs before the selection of the ASTRI design as the baseline. In this respect, the electroless nickel [31] (composed of nickel - Ni - and 9% to 12% of phosphorus – P -) is often added as a cladding layer of a few tens of micrometers onto the aluminum-based surface because of its hard and – at the same time - amorphous microstructure, which makes easier the super-polishing process of the optical surface. This allows us to achieve very low micro-roughness levels on metallic surfaces (up to a fraction of nanometer root mean squared (RMS), see e.g. [32, 33]).

Both the ASTRI M1 segments and the MST segments are realized using an improved version of glass cold-slumping technology[**], which was developed in synergy between OAB and ML. This approach has been already successfully used for the two telescopes of the MAGIC experiment array operated at the Observatorio di Roque de Los Muchacos astronomical site (La Palma, Canary Islands) [21, 34, 35]. The main steps of the mirror production are shown in Fig. 1.

---

[**] In this context, however, it should be noted that part of the mirrors for MST are being realized also with alternative technologies.

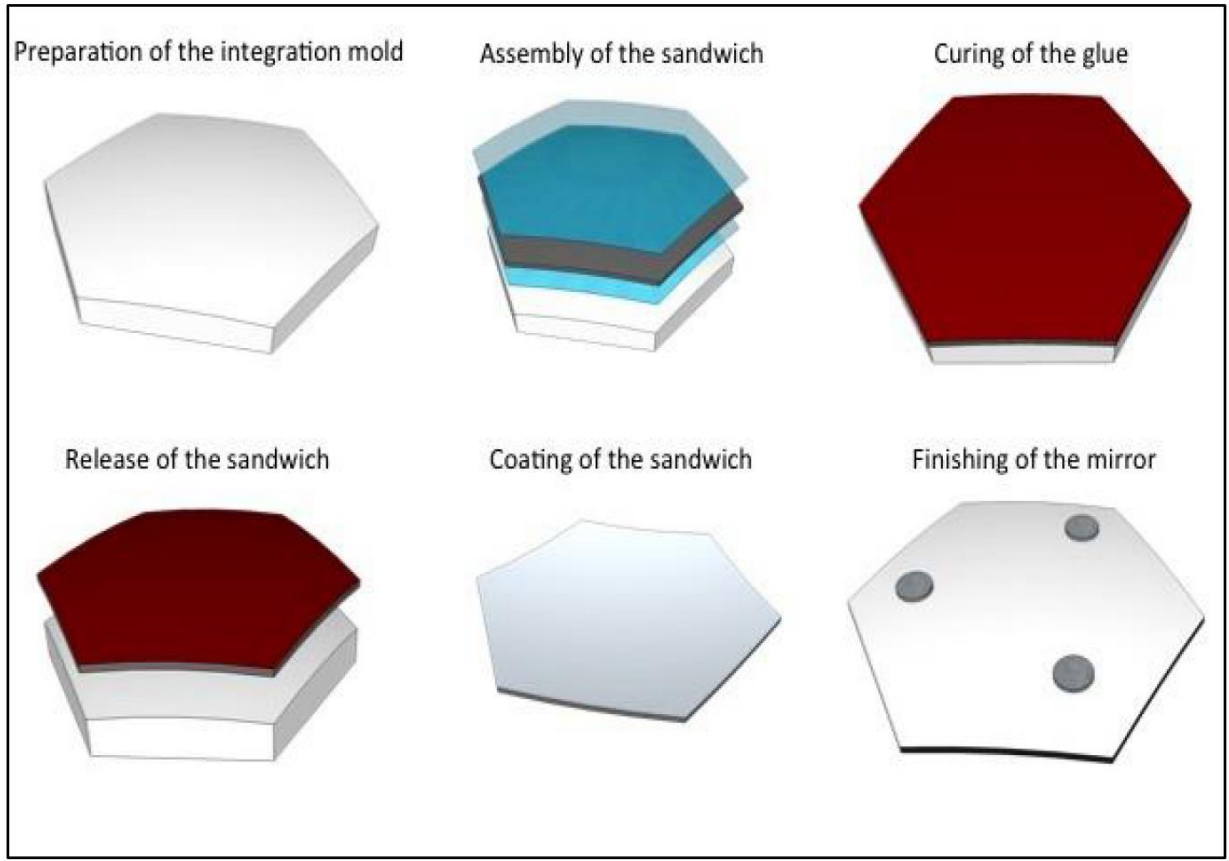

**Figure 1.** Conceptual description of the main steps (from top-left to bottom-right) of the cold-slumping technology.

The cold slumping is used to bend a thin glass foil onto a mould with the desired profile (Fig. 2a). Then, a honeycomb layer of aluminum (Fig. 2b) is glued onto the glass foil and, afterwards, a second glass foil is glued onto the honeycomb layer. In this way two glass foils are assembled in a light but stiff sandwich structure with an Al honeycomb core (Fig. 2c). After the curing of the glue, the sandwich is released and a highly reflective coating is applied to the outer surface of the inner glass foil (Fig. 2d). All the segments of the primary mirrors for the ASTRI MA and MST telescopes are realized using this technology, adapted to the two different cases [22].

In the case of the ASTRI M1 segments the thickness of the glass foils is 1.6 mm, while that of the honeycomb layer is 20 mm; for the MST segments the corresponding values are 2.1 mm and 30 mm, respectively. The reflective surface is a multilayer coating composed of $Al+SiO_2+ZrO_2+SiO_2$. The advantage of this technology is that it is based on a cost-saving replica process, which is suitable for CTA due to the multiplicity of the telescope mirrors. The chemical composition of the reflective layer guarantees a reflectivity profile higher than 85% in the wavelength range between 300 and 550 nm. Moreover, the evaporation process used to apply the reflective coating onto the mirror ensures a high reflectivity uniformity.

The segments obtained are lightweight (about 10 kg/m$^2$, excluding the three supporting pads, made in stainless steel, that increase the mass of less than two additional kilograms) and with high surface accuracy, which is measured with two different parameters, the micro-roughness and the residual shape error, both contributing in worsening the Point Spread Function (PSF). Both parameters are expressed as the RMS deviation of the reflecting surface from its ideal shape in the normal direction: but in the case of the micro-roughness the deviation is measured on microscopic spatial scales (with spatial wavelengths from a few micrometers up to 1 mm), at high spatial frequencies [36], while in the case of the shape error the deviation is measured at longer spatial scales. Surface micro-roughness affects the optical performance by degrading the PSF and reducing the image contrast [37], while the residual shape error causes deviation from the ideal focusing directions due to the geometrical imperfections. In accordance to the study performed by Taybaly et al. 2015 [36], the micro-roughness shall be less than 7.0 nm RMS, as measured in the 7-500 μm spatial wavelength range, in order to avoid spread of scattered photons > 15 % beyond a diameter of 0.12° of the focal spot.

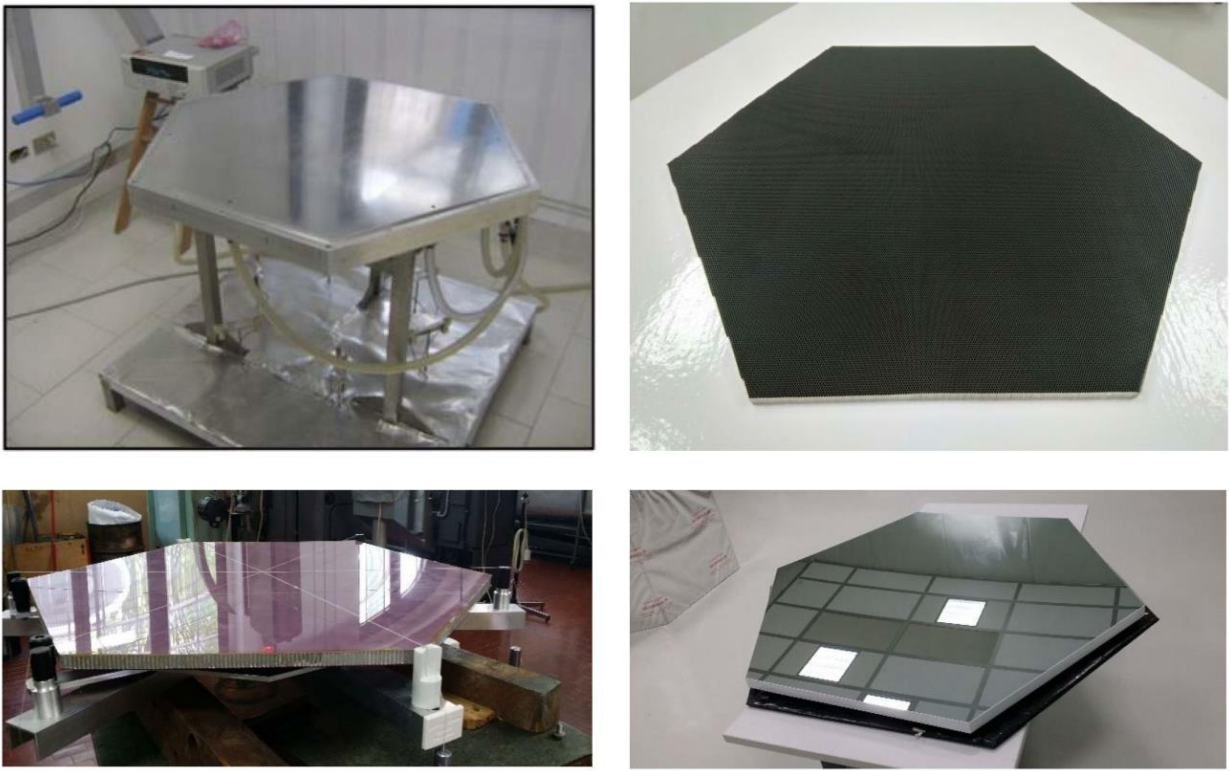

**Figure 2.** Pictures of the main steps of the mirror production with the cold-slumping method: bending of a thin glass foil onto the reference mould (*upper left*); honeycomb layer of aluminum to be glued onto the glass foil (*upper right*); sandwich structure of two glass foils with an Al honeycomb core (*lower left*); final mirror, with the reflective coating applied onto the first glass foil (*lower right*).

The accuracy of the mirror shape much depends on the accuracy of the reference moulds used for replication process, which have a residual shape error < 5 μm RMS compared with the expected shape. Since the replica process introduces a deviation lower than a factor 5 from the mould shape, the produced mirrors are characterized by a residual shape error < 25 μm RMS, that is well within usual requirements for the CTA telescopes.

Regarding the ASTRI M2 mirror, the system is based on a monolithic glass substrate with a thickness of 19 mm. This solution allows us to simplify the telescope design but requires a thick glass substrate, which cannot be curved to the required RoC with the cold-slumping technology used for the primary segments. Therefore, the ASTRI M2 mirror is bent to the correct shape using the hot-slumping technology. The reflective surface is obtained with the application of the same multilayer coating used for the ASTRI M1 and MST segments. Since the ASTRI M2 mirrors are produced with a completely different technology, the relevant production and qualification activities will not be described in this paper.

As mentioned in the Introduction (Section 1), the CTA southern array could be possibly extended after the implementation of the baseline configuration. In this context, additional Medium Size - 9 m diameter - aplanatic telescopes may be added with dual-mirror Schwarzschild-Couder configuration [7]. This design is capable to minimize the spherical and comatic aberrations across a large Field of View together with a reduction of the astigmatism by adopting a bended focal plane detection surface. An end-to-end prototype (pSCT, prototype Schwarzschild-Couder Telescope) has been successfully developed in US, Arizona [16], at the site at the Fred Lawrence Whipple Observatory. The telescope optical system of pSCT consists of a 9.7 m diameter primary mirror and a 5.4 m secondary. The primary is based on 48 individual mirror asymmetric petals: an inner and an outer rings of 16 and 32 asymmetric petals respectively (with an average area of 1.2 $m^2$). The secondary reflector is also segmented, with 8 petals forming an inner ring and 16 petals forming the outer ring (with an average area of 0.94 $m^2$). It should be noted that the reflecting panels have been realized by ML, supported also by INAF, using a glass replication technology almost identical to the production method adopted for the mirrors of ASTRI MA e MST. However, in the case of the innermost panels of the secondary mirror, due to the large sag of the segments, thin thermally pre-shaped glass foils have been used [38], instead of adopting a pure cold replication method. The use of pre-shaped glass foils via hot slumping for making the sandwiched panels presented a few challenging aspects, in particular related to the formation of micro-cracks on the surface of the glass that made the material less resistant to the stresses generated during the replication process. The pSCT telescope has been successfully tested and calibrated. Its functionality has been demonstrated with the characterization of the imaging capability [39] and the detection of the Crab nebula in gamma rays [40].

A couple of other cold replication approaches, and alternative to the one proposed by INAF and ML, have been proposed for making the reflecting segments of the MST telescopes. In particular, the solution based on open-structure composite segmented mirrors has been investigated by the Institute of Nuclear Physics of the Polish Academy of Sciences (IFJ PAN) [41]. Another similar method, but based on a close sealed structure, is the one investigated at IRFU-Saclay (France), with the panels made of layers of different materials [42]. At the time being, the two groups are engaged in an effort aiming at merging together the competences developed by the two groups, in order to deliver together all the reflecting panels for the 14 MST telescopes of the CTAO Southern Array.

In the case of the CTA LST telescopes, the mirrors are manufactured using the cold slumping technique and have a sandwich structure consisting of an aluminum honeycomb between two glass sheets, with a technology pretty similar to the one developed by INAF and ML [43, 44]. The parabolic primary mirror of CTA-LST is 23 m diameter and its focal length is 28 m. The primary mirror consists of 198 segmented mirrors. Each mirror has a hexagonal shape of 1.51 m side-by-side size. All the mirrors for the first LST have been already produced and successfully installed.

It should be noted that Will et al. (2019) [45], in collaboration with ML, are testing a new kind of back-coated glass mirrors for the MAGIC Telescopes. This technology represents an evolution of the INAF glass replication approach, but in this case a protective layer of an ultra-thin glass sheet which is back-coated with aluminum is used instead of the $SiO_2$ thin coating deposited under vacuum onto the aluminum reflecting layer. Finally, another promising replication process under investigation by INAF and ML foresees a new fabrication scheme, similar to the cold glass replication, but where the reflecting slab is a low-cost laminate pre-coated reflecting aluminum strips instead of glass sheets[††] [46].

## 4 Applicable Requirements

For the development of the manufacturing process for the mirror segments, the CTA requirements have been assumed, not only for the case of the MST tiles but also for the one of the ASTRI MA (which are the same of the SST Telescopes). These reference parameters regard not only the optical performance of each item but also its reliability, taking also into account the environmental conditions of the observing sites (Table 1) and the durability prescriptions.

**Table 1.** Applicable CTA environmental requirements

| REQUIREMENT DESCRIPTION | VALUE |
|---|---|
| Scientific observations shall be possible when air temperature is: | -15 °C < T < +25 °C |
| Telescopes shall suffer no damage when air temperature is: | -20 °C < T < +40 °C |
| Telescopes shall suffer no damage when air temperature gradient at night time is: | < 7.5 °C/h |
| Telescopes shall suffer no damage when air temperature variations within 24 hours are: | ± 30 °C |
| Telescopes shall withstand continuous air temperature gradients of at least: | 0.72 °C/h for 25 min |
| Scientific observations shall be possible when humidity is: | 2-90 % |
| Telescopes shall suffer no damage when humidity is: | 2-100 % |
| Telescopes shall suffer no damage when rain in 24 hours is: | < 200 mm |
| Telescopes shall suffer no damage when rain in 1 hour is: | < 70 mm |
| Telescopes shall suffer no damage for precipitation in the form of rain, snow and hail when wind speed is: | < 90 km/h |
| Telescopes shall suffer no damage during slewing to safe position when rain is: | < 0.5 mm/min |
| Telescopes shall suffer no damage when snow accumulation on ground is: | < 50 cm |
| Telescopes shall suffer no damage by the impact of hailstones with diameter: | < 20 mm |
| Telescopes shall suffer no damage when the ice thickness (on all surfaces) is: | < 20 mm |
| Scientific observations shall be possible when wind speed is: | < 36 km/h |
| Telescopes shall suffer no damage during slewing to safe position when wind speed is: | < 50 km/h |
| Telescopes shall suffer no damage when the 10-minute average wind speed is: | < 120 km/h |
| Telescopes shall suffer no damage when the 1-second average wind gusts are: | < 200 km/h |

From the performance point of view, the initial reflectivity of all facets, at any wavelength in the reference range 300-550 nm, must be > 85 %. This requirement should be fulfilled also for the secondary mirrors of the ASTRI MA (and therefore CTA SST) telescopes. In this respect, the rays, after the first reflection by the primary, impinge the secondary mirrors with angles between 10º and 50º with respect to the optical axis. On the other hand, the theoretical reflectivity of Aluminum coated mirrors with $SiO_2$ protection (utilized for the monolithic secondary mirrors and deposited via evaporation [47]) remains very high within this entire angular range; experimental verifications were also carried out on witness samples.

In addition, there is a requirement on the angular size of the telescope PSF. To this aim, the reference parameter is $\theta_{80}$, which is the angular diameter of the circle that includes the 80 % of the focused photons. For SST (and therefore also ASTRI MA) and MST telescopes, the CTA requirement is $\theta_{80} < 0.25°$ and $\theta_{80} < 0.18°$, respectively.

---

[††] G. Pareschi, et al., Patent Application number: 102021000019658 submitted on 23 July 2021 to the Ministero dello Sviluppo Economico of the Italian Government with the title «Metodo per la costruzione di un elemento ottico per un telescopio ed elemento ottico ottenuto con tale metodo» (in English: *Method for the construction of an optical element for a telescope and optical element obtained with such a method*)

It should be noted that the PSF requirement for the MST is strictly referred just to the on-axis direction, in order to allow the quality assurance verification of each spherical tile. Indeed, while the SST (and ASTRI MA) telescopes are aplanatic within their FoV, one should be reminded that the MST optical system is instead affected by an inherent off-axis degradation of the PSF, as they are based on the single-mirror Davis-Cotton design [29]. A worsening of the angular resolution is then associated to the increase of the off-set angles and the PSF also becomes asymmetric in the two directions, with the spot size which is always larger in the radial direction. This is mostly due to the non-Gaussian tails of the PSF and the effect tends to be more enhanced with the increase of the focal length/diameter ratio. At the border of the FoV of the MST telescopes (± 3.7°), the maximum degradation due to the design will be about 0.10°, i.e. in any case lower than the angular resolution requirement. Therefore, also for off-axis rays, the PSF of the MST is expected to be dominated by the fabrication and alignment errors.

The main environmental requirements are reported in Table 1; furthermore, there are also requirements on the resistance to aggressive atmosphere, dust and sand blasting, and lightning. Finally, the provided mirrors and segments must have a lifetime > 15 yr. Of course, this last parameter could not be directly proven within the activities and tests performed and reported in the present work. However, the experience coming from the glass sandwiched mirrors on the MAGIC II telescope since 2007, that up to now have shown a very small degradation, and the aging tests performed so far indirectly indicate that this requirement can be met.

In view of the mirror production, all the CTA environmental requirements reported in Table 1 were considered as applicable requirements for ML, the industrial supplier of MST and ASTRI M1 segments; we only reinforced the air temperature range where telescopes shall suffer no damage (-20 °C < T < +70 °C), to take into account also the transport and storage conditions before the segment integration on the telescope. Moreover, we required that the mirror segments are water-proof and that the reflective coating can survive to a tape adhesion test with a pulling force up to 16 N. Finally, from the reliability point of view, we required that the lifetime of the reflective coating is > 6 yr, with an end-of-life reflectivity > 65 % in the reference wavelength range. A number of aging tests on the coating were performed [23], while one should note that the reflectivity loss of the mirrors mounted on MAGIC II at the Observatorio del Roque de los Muchachos (i.e. the same astronomical site in La Palma, Canary islands, where the northern array of CTA will be operated) was less than 2 % after more than 5 years[‡‡]. Recent tests performed on a MST mirror that was mounted on a truss in Paranal (Chile) at the site where the CTA southern array will be installed, didn't show any degradation after 1 year of exposition to the local environment[§§].

From the performance point of view, we required an initial reflectivity value > 85 % in the wavelength range 300-550 nm. In addition, we required a maximum value < 8 % (over the entire mirror surface) for the reflectivity non-uniformity, which is defined as the maximum difference between the average surface reflectivity measured at different positions over the mirror.

Concerning the micro-roughness, the criterion adopted for the production was to maintain the already excellent RMS level <2 nm, i.e. a value much lower than the maximum acceptable value of 7 nm RMS estimated by Tayabaly et al (2015, 2016) [36,48] in order to not exceeding the MST $\theta_{80}$ requirement because of scattering.

---

[‡‡] Markus Garczarczyk, Reflectivity of the MAGIC-II reflector, MAGIC Document 111003 (2011)
[§§] M. Garczarczyk, Report from the Chile mirror test campaign, CTAO MST-STR-RD-36144000-00011-2 Issue 1 (2021)

We required that the micro-roughness be < 2 nm RMS, thus constraining the dispersion of peaks and valleys of the real surface profile. For the MST segments we required a $D_{80}$ (the linear dimension of $\theta_{80}$) < 12 mm at the nominal focal length for the on-axis single mirror, which is half of the RoC (due to the spherical shape of the mirror). For a focal length of 16.07 m this corresponds to about 0.045°, which is well within the CTA requirement. Regarding the radius of curvature, we considered as a guideline that RoC = 32.14 ± 0.04 m. For the ASTRI M1 segments there is no requirement on the RoC, since they are aspheric. In their case we introduced two requirements, one on the radius of the spherical component (between 8198 and 8298 mm) and one on the residual shape error (< 20 μm RMS for corona 1 and < 25 μm RMS for coronae 2 and 3). Finally, for both the MST and ASTRI M1 segments the tolerance on the hexagon side-to-side size is ± 2 mm.

## 5 Qualification of the Production Process

Before starting with the mass production of the mirror segments, it was necessary to validate the production process and to perform qualification tests on a set of preliminary segments. To this end, the industrial activity started in February 2017 with the development phase. During this phase, the moulds used to produce the different types of segments were qualified and the parameters of the glass cold-shaping process were fine-tuned, in order to satisfy the performance requirements of the mirrors. Several segment prototypes were then produced and tested along this period, in order to ascertain that the production process was stable and repeatable. Afterwards, one qualification segment of each type went through a complete set of performance measurements and environmental and reliability tests. During the qualification process, the following environmental tests were performed on each qualification segment:

1) 5 thermal cycles composed of the following steps:
- From 20°C to -20°C
- Plateau of 12 h
- From -20°C to 70°C
- Plateau of 4 h
- Cool down to 20°C

For all steps the temperature gradient is 0.125°C/min
2) Dry Heat test (according to ISO 9022-2:2002(E))
3) Dump Heat test (according to ISO 9022-2:2002(E))

Before and after each of the above tests, the tested segment was characterized in terms of integrity, reflectivity, and residual shape error map; this was computed as difference between the measured shape, verified with a POLI TCX Coordinate Measurement Machine (CMM) on a 10 mm square grid, and the designed one. In all cases, after the test: 1) the visual inspection revealed no degradation in term of panel integrity and quality; 2) the measured reflectivity was fully compatible with the original one; 3) the comparison of the residual maps revealed no significant variation of the surface shape error. Moreover, the immersion in a water tank was executed, in order to verify the water tightness.

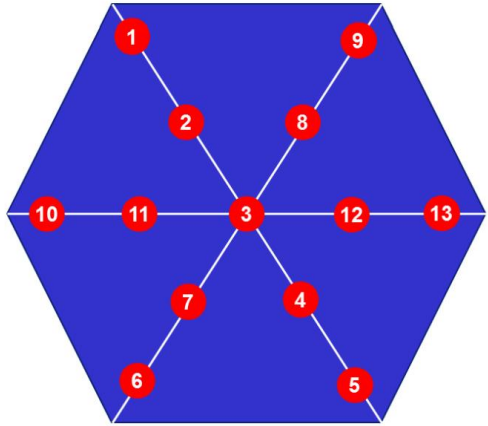

**Figure 3:** Distribution of the 13 points used to qualify the mirror reflectivity

Regarding the reflecting coating, the reflectivity curve on the wavelength range 300-550 nm was measured on 13 equidistant points along three diagonals (Fig. 3), in order to sample peripheral, intermediate, and central areas of the mirror. The adhesion of the mirror coating was verified through a tape removal test. Both the reflectivity measurements and the tape test were repeated after the execution of thermal cycles, dry- and damp-heat tests, and solar radiation tests. All these measurements and tests demonstrated that the performance of the produced segments are compliant with the applicable specifications and that the segments do not degrade due to the environmental conditions.

Once the production process was properly qualified, the serial mirror production was approved. It was performed following the production flow reported in Fig. 4.

## 6  Quality Assurance Approach

Once the production process was qualified, in February 2018 it was possible to start with the mirror mass production. In order to further consolidate the manufacturing and verification flow chart of Fig. 4, it has been decided to consider these first sets of panels as a pilot production, and to follow a "first article inspection" approach (cf. EN 9102). The panels have been subdued to an extensive set of acceptance verifications at the manufacturer site, double-checked independently for the most critical parameters at the Customers laboratories. The verification approach and the relevant tests for the different sets of mirrors are described in the present Chapter 6; the obtained results are detailed in Chapter 7.

First of all, for each mirror segment a series of acceptance tests and measurements were performed by ML. As a first step, a visual inspection was executed, to ascertain the mirror integrity and evaluate the cosmetic defects (such as scratches, spots, and halos) of the reflective surface. Then, a tape removal test was performed, in order to verify the coating adhesion. Afterwards, different types of measurements were performed in order to characterize the optical and metrological properties of each segment.

The reflectivity curves were measured in five different equidistant points along one diagonal: two positions are near two opposite segment vertexes and one is at the segment center, while the remaining two positions are at intermediate distance between the center and the two opposite vertexes. In this way, the five reflectivity-curves provide a representative picture of the whole segment. The average reflectivity and the reflective non-uniformity were derived from these curves.

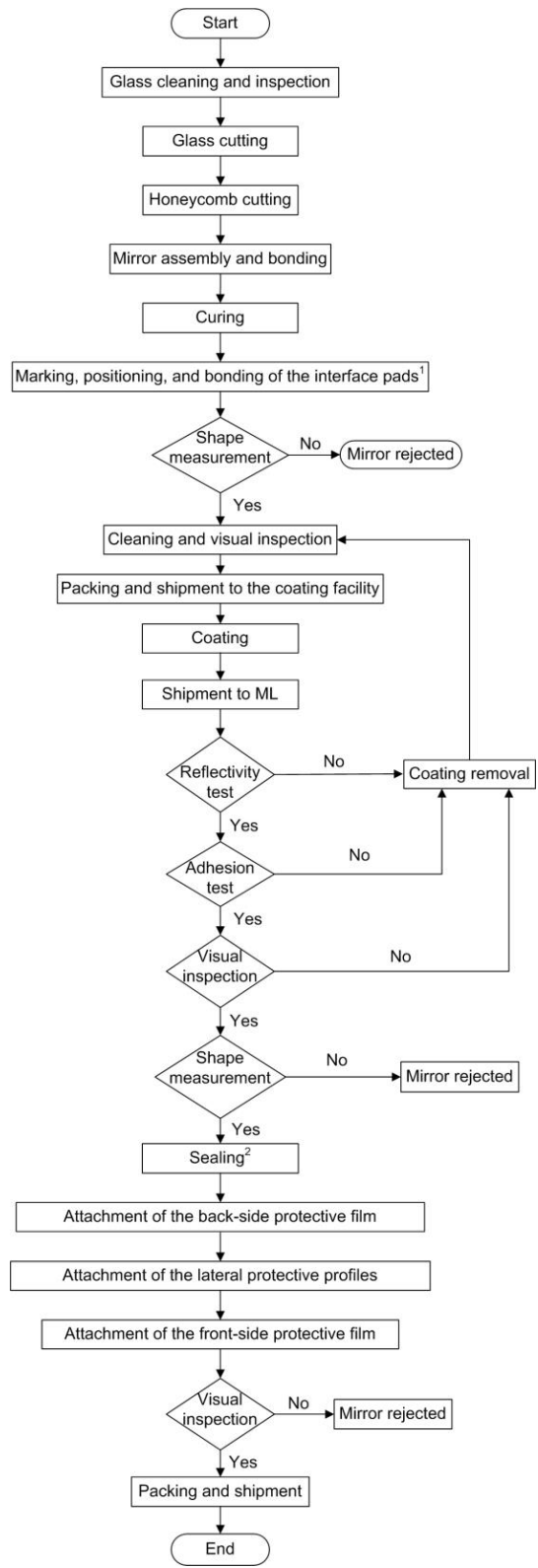

**Figure 4.** Production flow for the ASTRI M1 and MST mirrors. Notes: 1) Three pads are installed on each mirror to provide mechanical interface with the telescope mechanical structure; 2) The lateral surfaces of each mirror are sealed with silicone to prevent the penetration of atmospheric agents

From the metrological point of view, in the case of the focusing MST segments, the measurement of the RoC and of the PSF size (at both the best and the nominal focal length) was performed. In the case of the aspheric ASTRI M1 segments, instead, the best-fit value of the spherical component and the residual shape error were measured. The accuracy of the RoC measurement performed with the CMM is about 50 mm and is obtained by spherical best-fit of the measured data points. In production, the optical surface is sampled in 536 points. The above uncertainty has been estimated by comparison with the shape of the mirror derived from measurements done with a much finer grid, which is impractical in an industrial environment.

The measurement of the PSF size is obtained with an optical bench, based on a $2f$ configuration, which generates a spherical wave and detects the focused back-reflected light beam. The set-up of this facility consists of a point-like (with respect to mirror dimension and curvature) laser source at 632 nm that is converted into a spherical wave by means of a microscope eye-piece. The required numerical aperture is about 0.016 and it is easily achieved. The MST mirror is positioned at a distance of 32140 mm with an accuracy of the order of 1 mm. The back-reflected radiation illuminates a screen positioned at the focal plane, which is in turn imaged on a 2048 × 2048 pixel CCD camera. The camera covers a physical area of 170 mm × 170 mm. In order to avoid superposition between the source and the image plane, each of them is transversally displaced from the optical axis by 100 mm. The measurement of the $D_{80}$ diameter has an accuracy of 3 mm. This is due to several reasons, which include the contribution of variable daylight illumination (despite of the presence of a filter in front of the CCD), the finite area of the focal plane imaged onto the CCD, the scattering of light from the screen. The entity of the accuracy (3 mm) has been estimated by repeated measurements and comparison with controlled laboratory conditions, which of course are not completely possible in an industrial environment.

Apart from the acceptance tests performed on each segment, additional tests were performed by ML on some individual segments regarding the micro-roughness and the water tightness. These tests were necessary in order to keep under control the production process.

The contract assigned to ML requires that each mirror segment is associated with an "Identity Card" (IC), which reports the results of all the acceptance tests and measurements. The IC is the reference document to control the mirror quality. Along the production phase, the Quality Assurance team used it to monitor the production process and decide which segments required a direct inspection, in order to evaluate if they can be accepted or rejected. In case a segment showed a major anomaly (see Fig. 5), which however did not affect its performance or reliability (such as a scratch or a wide halo), it was accepted only if a comprehensive description of the anomaly was reported in a Non-Conformance Report (NCR). If, on the other hand, the detected anomaly compromised the segment performance or reliability (such as a large spalling of the mirror edge), the segment was rejected. After its delivery, each segment must be followed by its IC (and the relevant NCRs, if any) for all its life. In this way, at any time it will be possible to retrieve the mirror properties and evaluate if it is suitable for the installation on the telescope.

Additional sample tests were performed by OAB and DESY at their laboratories on ~ 10% of the produced mirrors. Such tests concerned the RoC and PSF measurement for the MST segments and the shape measurement for the ASTRI M1 segments. In this way we obtained an independent check of the segment properties, monitored the reliability of the production process and provided feedback to the industrial supplier.

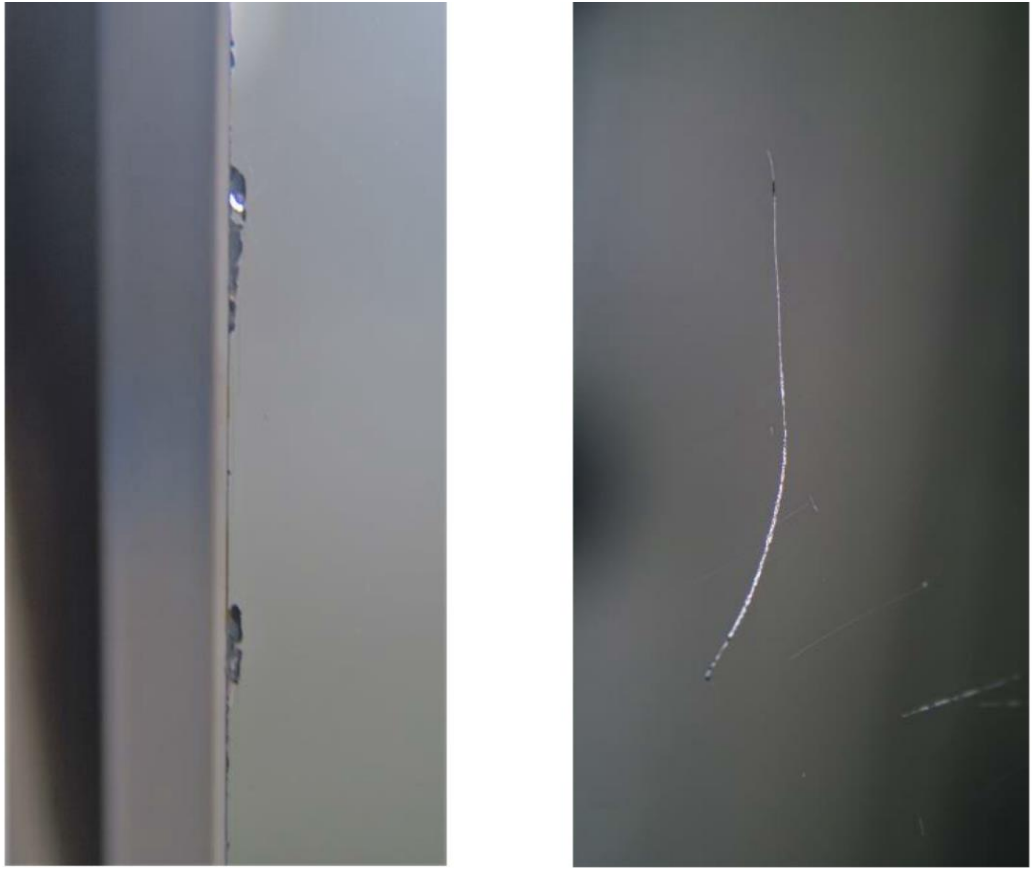

**Figure 5.** Examples of the most common mirror defects: edge irregularities (*left*) and coating scratches (*right*)

The 200 MST segments were delivered by the industrial supplier in 4 different batches of ~ 50 segments each. For each of these batches, we selected 5 segments (~ 10 % of the whole batch) for additional measurements. In all cases, the 5 segments were selected as follows: 2 among the segments with the best PSF (i.e. small $D_{80}$); 2 among the segments with the worst PSF (i.e. large $D_{80}$); 1 segment with intermediate PSF. We made such selection in order to check whether the results obtained at ML and those obtained with additional measurements depend on the mirror shape.

For 15 of these MST segments the additional measurements were performed at INAF-OAB, using an outdoor optical bench based on a $2f$ configuration, installed at the laboratories of the Brera Astronomical Observatory in Merate (Italy) [49]. Other 5 MST segments (different from those measured at OAB) were instead verified at the DESY $2f$ optical bench [50]. In both cases the $2f$ facility uses a LED as light source (without any diffuser, since the generated wave-front is already spherical) and a CCD camera as image detector: in this way a photometric image is obtained, which is characterized by high homogeneity and uniformity.

For each tested MST segment the following measurements were performed:
- Mirror RoC, performed by acquiring a series of images of the focal spot along the optical axis
- Mirror $D_{80}$ at both the nominal focal length (NF), corresponding to the nominal $2f$ value of 32.14 m, and the best focal length (BF), which is the $2f$ value where the $D_{80}$ value is minimum

The final measurement regarded the focused reflectivity of the MST segments. In fact, while the average reflectivity reported in the IC of each segment is the average of the surface reflectivity measured at five different positions along one diagonal, the focused reflectivity is the reflectivity of the overall segment measured at the focal plane. Therefore, it is a parameter which truly represents the CTA requirement. The measurement of this parameter has been performed at the *2f* UV-CCD facility of the Institut für Astronomie und Astrophysik (IAAT) of the University of Tübingen (Germany) [51], using the same five MST segments which were verified at DESY.

## 7 Results

### 7.1 Overview

The mirror production was completed at the end of 2019. The whole set includes 200 MST segments and 200 ASTRI M1 segments (68 COR1, 66 COR2, and 66 COR3). The MST segments correspond to two full mirrors (of 86 segments each) plus 28 spare segments. In the case of ASTRI M1, the delivered segments correspond to 11 complete M1 mirrors (of 6 × 3 segments each) plus two additional COR1 segments for test purposes.

### 7.2 Results for MST segments

#### 7.2.1 Results obtained at ML

In Fig. 6 we show the main parameters of the MST segments, as reported in the relevant Identity Cards delivered by the industrial supplier. In the first panel the reference nominal value is represented with a green solid line, while in the other panels the applicable requirement value is represented with a red solid line. In all panels the average value is represented with a black dashed line; the colored regions above and below the average value correspond to 1, 2, and 3 standard deviations ($\sigma$) from the average value, respectively.

The upper panel reports the measured value of the RoC. For almost all the segments, this value is close to the nominal value of 32.14 m (although in many cases it is not within the range 32.10-32.18 m) and shows no tendency to increase or decrease. The only exceptions regard the first two segments, which have a significantly smaller radius. They were produced when the process was not yet completely assessed. This means that, after these first two segments, the production process has remained stable. In any case, it should be noted that the requirement on the RoC is not considered an applicable acceptance criterion, since what matters is only the value of $\theta_{80}$ at the nominal focal length. On the other hand, for all but one of the segments produced so far, the $D_{80}$ value was lower than 24 mm at twice the focal length (@ *2f*, second panel): this implies that these mirrors are considered compliant with the CTA requirement of $D_{80} < 12$ mm @ *1f*. Moreover, apart from this isolated worst case, the measured data show no long-term increasing or decreasing trend in the production. The average reflectivity (third panel) is well above 85 % for all the segments, and the average reflectivity non-uniformity of the mirrors (bottom panel) is always below the requirement of 8 %. Therefore, also these parameters are compliant with the corresponding input requirements.

In Fig. 7 the previous results are summarized in histograms, which report the distribution of each parameter around its average value. They show that the distribution is rather symmetric for the RoC and the reflectivity, while it has a tail towards the high values for $D_{80}$ and the reflectivity non-uniformity.

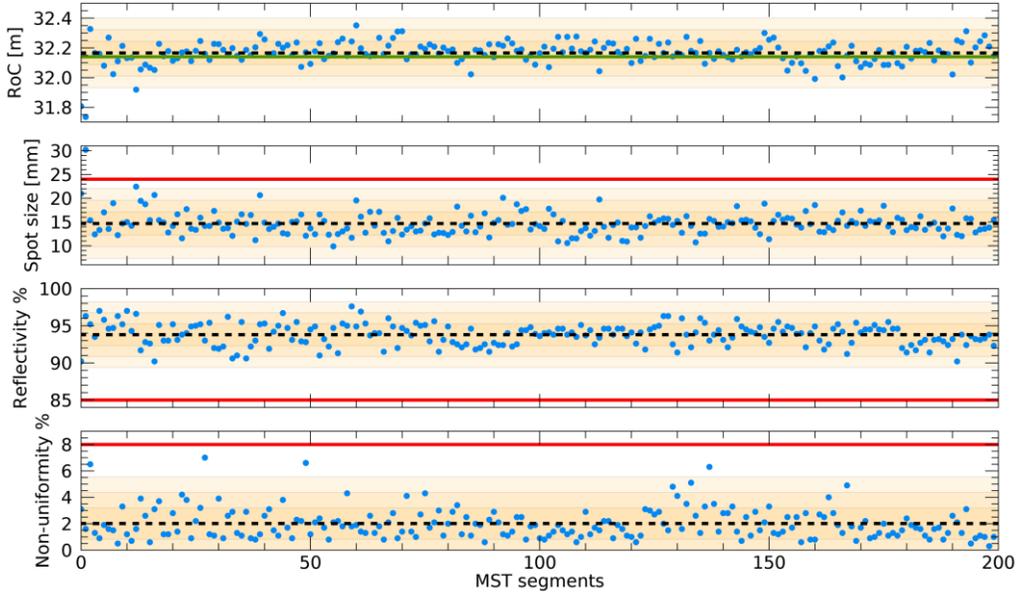

**Figure 6.** Main parameters of the produced MST segments, from the first (segment 0) to the last made (segment 199): radius of curvature (*upper panel*), D$_{80}$ at twice the nominal focal length (*second panel*), average reflectivity (*third panel*), and reflectivity non-uniformity (*bottom panel*). In the first panel the green solid line represents the reference nominal value, while in the other panels the red solid line represents the applicable requirement value. In all panels the black dashed line represents the average value.

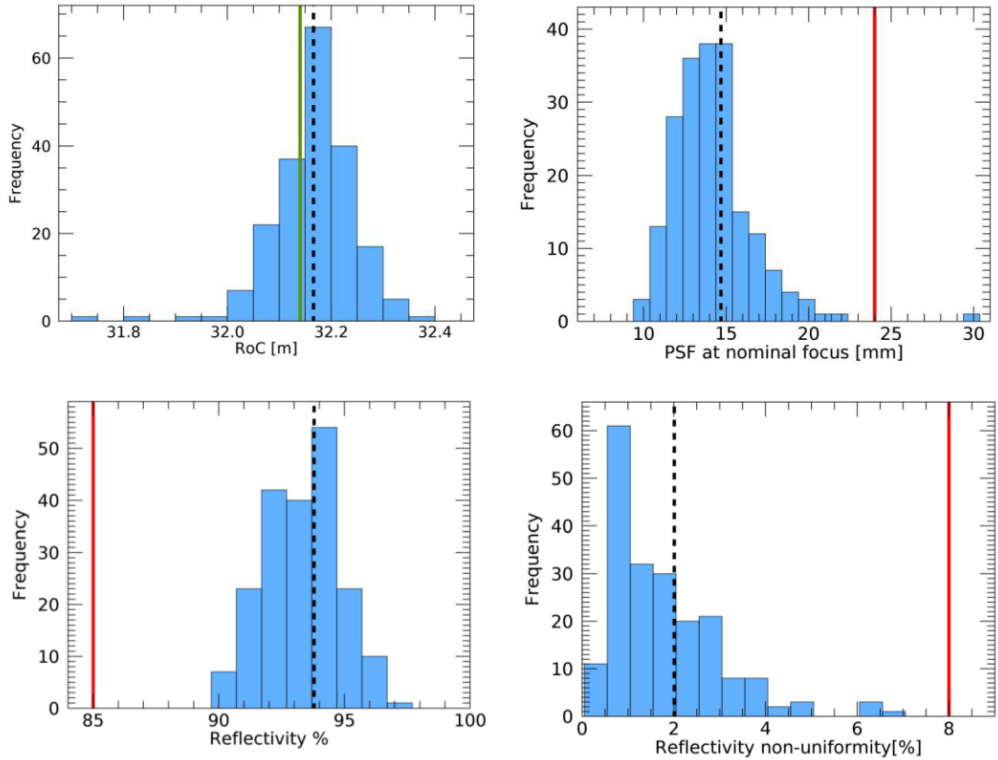

**Figure 7.** Distribution of the main parameters for the MST segments: radius of curvature (*upper left*), D$_{80}$ at twice the nominal focal length (*upper right*), average reflectivity (*lower left*), and reflectivity non-uniformity (*lower right*). In the upper left panel the green solid line represents the reference nominal value, while in the other panels the red solid line represents the applicable requirement value. In all panels the black dashed line represents the average value.

*7.2.2 Results of sample tests performed at INAF-OAB and DESY*

In Fig. 8 we report, for each of the 20 sample segments which were independently double-checked, the difference between the RoC values measured at the CMM in ML (those reported in the segment ICs) and the corresponding values measured at either the OAB or the DESY $2f$ facility. It shows that, while there is no systematic difference between the results obtained at OAB and DESY, for most segments the RoC measured by ML is larger than that obtained with the other facilities. This difference is likely due to the different segment position during the measurement: in fact, the segment lies horizontally during the measurement with the ML CMM, while it is mounted vertically at the $2f$ facility. This implies an opposite impact of the gravity, which increases the RoC in the first case and reduces it in the second one.

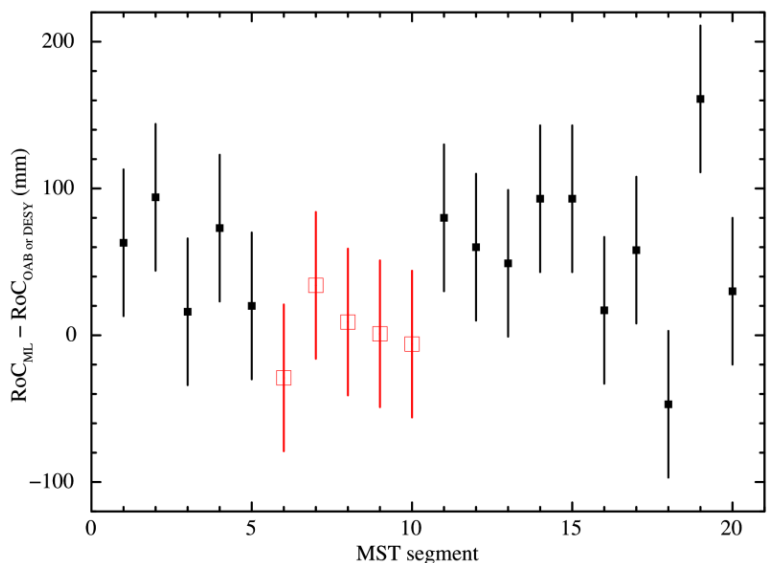

**Figure 8**. Difference between the mirror Radius of Curvature measured at ML and either OAB (filled black squares) or DESY (empty red squares). The error bars indicate the systematic error of the CMM.

Taking into account these points, the values measured at OAB or DESY are compatible with those provided by ML.

In Fig. 9 we report, for each segment, the difference between the $D_{80}$ measured by ML and the corresponding value measured at either the OAB or the DESY $2f$ facility, at both best and nominal focal length. It shows that also in this case the results obtained by OAB and DESY are compatible. On the other hand, at both the BF and NF position, for most segments the $D_{80}$ value obtained at OAB or DESY is slightly lower than the corresponding value obtained by ML. This difference is due to the different methods used by ML and OAB or DESY to perform the $D_{80}$ measurement. In fact, the measurements performed at OAB and DESY use a LED as light source (without any diffuser, since the generated wave-front is already spherical) and a CCD camera as image detector: in this way a photometric image is obtained, which is characterized by high homogeneity and uniformity.

These differences in the measurement approach explains the larger $D_{80}$ value obtained (on average) by ML, compared with that obtained at OAB and DESY. Taking into account the intrinsic uncertainty of these measurements, the two methods provide results that can be considered compatible; the results just confirm that $D_{80}$ is well within the specification for all segments also using a different configuration for the characterization.

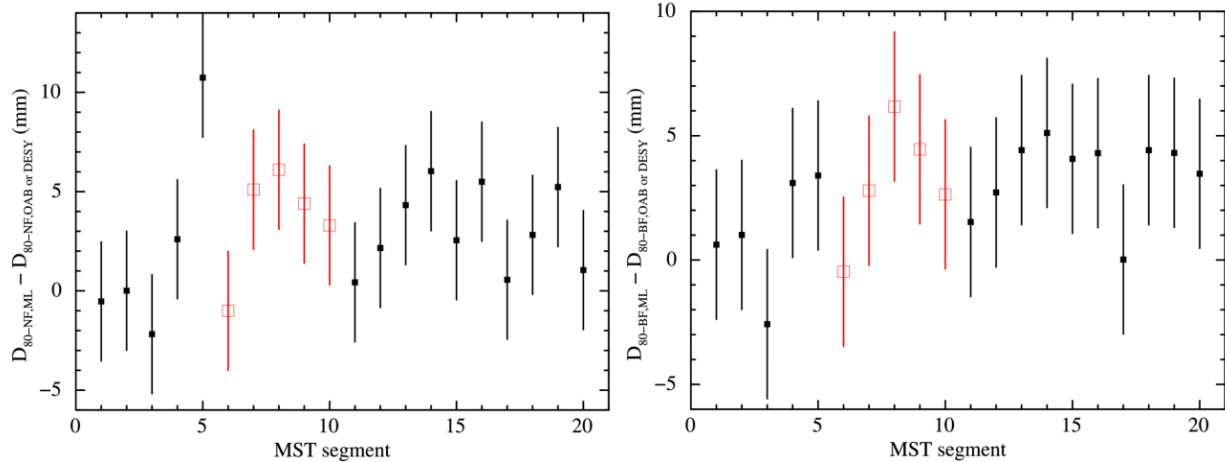

**Figure 9.** Difference of the $D_{80}$ measurements at ML and either OAB (filled black squares) or DESY (empty red squares), at nominal focus (*left panel*) and best focus (*right panel*). The error bars indicate the systematic error of the ML measurements.

### 7.2.3 Results of sample tests performed at IAAT

The *2f* principle assumes a point-like light source, which is realized at IAAT with LEDs of different wavelengths (310 nm, 365 nm, 405 nm, 490 nm, 525 nm) driven by a highly stable current supply. A spherical mirror placed at a distance equal to its RoC, i.e. two times its focal length *f*, generates an image of the light source at the same distance. We project this image on a diffuse polytetrafluoroethylene (PTFE) screen and record it with a UV-sensitive CCD camera.

The focused reflectivity of the MST segments was determined by integrating the reflected intensity over a circular area with 2 mrad radius around the center of the spot and dividing by the incident intensity on the mirror. The incident intensity was measured by placing the screen between light source and mirror and integrating over the projected mirror area. Moreover, the same data were used to determine the mirror PSF, which was then averaged over the five wavelengths. 100 % of the intensity for the $D_{80}$ determination was defined as the intensity in a circular area with 2 mrad radius around the center of the spot. This angular scale refers to the *2f* distance of the measurement, thus, in the MST case with a nominal RoC of 32.14 m, 2 mrad correspond to a radius of about 64 mm.

In Table 2 we report, for each of the five MST segments, the focused reflectivity at each of the five wavelengths and the average value of $D_{80}$. It shows that, for all mirrors, the reflectivity reaches 85 % at all wavelengths and the $D_{80}$ value is ≤ 14 mm, in agreement with the input requirements.

**Table 2.** Focused reflectivity and PSF spot size measured at IAAT

| Mirror | Reflectivity (%) | | | | | $D_{80}$ (mm) |
|---|---|---|---|---|---|---|
| | 310 nm | 365 nm | 405 nm | 490 nm | 525 nm | |
| MST-ML-114 | 85±2 | 88±2 | 89±2 | 86±2 | 86±2 | 12.5±0.2 |
| MST-ML-118 | 88±2 | 92±2 | 94±2 | 91±2 | 91±2 | 14.0±0.3 |
| MST-ML-122 | 88±2 | 90±2 | 91±2 | 88±2 | 88±2 | 13.4±0.2 |
| MST-ML-123 | 85±2 | 91±2 | 92±2 | 91±2 | 89±2 | 12.3±0.3 |
| MST-ML-124 | 85±2 | 90±2 | 91±2 | 88±2 | 88±2 | 14.0±0.1 |

### 7.2.4 Summary on MST results

In Table 3 we report the summary of the results obtained with the various metrology measurements performed on the MST segments. For all the sets of segments the average RoC is in the range 32.1-32.2 m, with a standard deviation of 5-10 cm around the average value. This means that, in all cases, the average RoC is very near the nominal value of 32.14 m, and that the dispersion around this value is very low. Regarding the PSF size, for all sets the average $D_{80}$ value is in the range 11-15 mm, with a standard deviation of 1-3 mm around the average value. Therefore, also in the case of the PSF size all measurements confirm that the produced segments are well within the requirement of 24 mm @ $2f$.

**Table 3.** Summary of metrology results obtained for the MST segments.

| Laboratory | Measured parameter | Measuring tool | Number of measured segments | Results |
|---|---|---|---|---|
| Media Lario metrology | RoC | CMM 3D machine | 200 | Average value: 32166 mm<br>Standard deviation: 78 mm |
| Media Lario optical | D80 @ twice nominal focal length | 2f facility<br>• point-like laser source<br>• microscope eye-piece<br>• CCD camera: 2048x2048 pixels | 200 | Average value: 14.7 mm<br>Standard deviation: 2.5 mm |
| OAB | RoC | 2f facility<br>• led source<br>• CCD camera: 2048x2048 pixels of 24x24 μm | 15 | Average value: 32106 mm<br>Standard deviation: 92 mm |
| OAB | D80 @ twice nominal focal length | 2f facility<br>• led source<br>• CCD camera: 2048x2048 pixels of 24x24 μm | 15 | Average value: 12.8 mm<br>Standard deviation: 2.9 mm |
| DESY Berlin | RoC | 2f facility<br>• led source<br>• CCD camera: 1360x1024 pixels of 4.65x4.65 μm | 5 | Average value: 32191 mm<br>Standard deviation: 55 mm |
| DESY Berlin | D80 @ twice nominal focal length | 2f facility<br>• led source<br>• CCD camera: 1360x1024 pixels of 4.65x4.65 μm | 5 | Average value: 11.1 mm<br>Standard deviation: 1.6 mm |
| IAAT Tubingen | D80 @ twice nominal focal length | 2f facility<br>• led source<br>• CCD camera: 1024x1024 pixels of 13x13 μm | 5 | Average value: 13.2 mm<br>Standard deviation: 0.8 mm |

In Table 4 we report instead the summary of the reflectivity measurements performed at ML (for the surface reflectivity) and at IAAT (for the focused reflectivity). It shows that, in both cases, the average reflectivity value is well above the requirement of 85 % and the standard deviation around it is below 2 %. This result confirms the high stability reached with the adopted production process.

**Table 4.** Summary of the reflectivity results obtained for the MST segments.

| Laboratory | Measured parameter | Measuring tool | Number of measured segments | Results |
| --- | --- | --- | --- | --- |
| Media Lario | Surface reflectivity | Measurement on five different points using a thin-film analyzer ($\lambda$ = 190-1700 nm) with a contact probe and Deuterium-Tungsten light source | 200 | Average value: 93.8 % <br> Standard deviation: 1.5 % |
| IAAT Tubingen | Focused reflectivity | 2f facility with a led source and a CCD camera | 5 | Average value: 89.0 % <br> Standard deviation: 1.6 % |

These results show that the measurements of the RoC and $D_{80}$ performed on the MST segments with different facilities provided compatible values and give us high confidence on both the reliability of the production process and the compliance of the mirrors with the CTA requirements.

### 7.3 Results for ASTRI M1 segments

Regarding the ASTRI M1 segments, in this section we report the results obtained for all the 68 COR1 segments, while in the Appendix we report the corresponding results for COR2 and COR3 segments. In all cases they are retrieved from the relevant Identity Cards delivered by ML.

In the first and second panel of Fig. 10 we show, respectively, the values of the RoC and the residual shape errors (we remind here that the ASTRI M1 segments are aspherical and that RoC refers to the spherical component). For all COR1 segments the first parameter is within the required range (between 8198 and 8298 mm) and the second parameter is < 20 µm RMS, in agreement with the corresponding requirement. Regarding the mirror reflectivity, in the third and fourth panel we report, respectively, its average value and its non-uniformity for each segment. As for the MST segments, also for all the ASTRI COR1 segments these parameters are compliant with the lower limit of 85 % and the upper limit of 8 %, respectively.

In Fig. 11 the previous results are summarized in histograms, which report the distribution of each parameter around its average value. They show that, for all the COR1 segments, the values of RoC and accuracy have a wide and rather uniform distribution around the average value, while the distribution of the reflectivity values is rather peaked around the average value. Regarding the reflectivity non-uniformity, most of the segments have a low value, but the overall distribution shows a tail at high values.

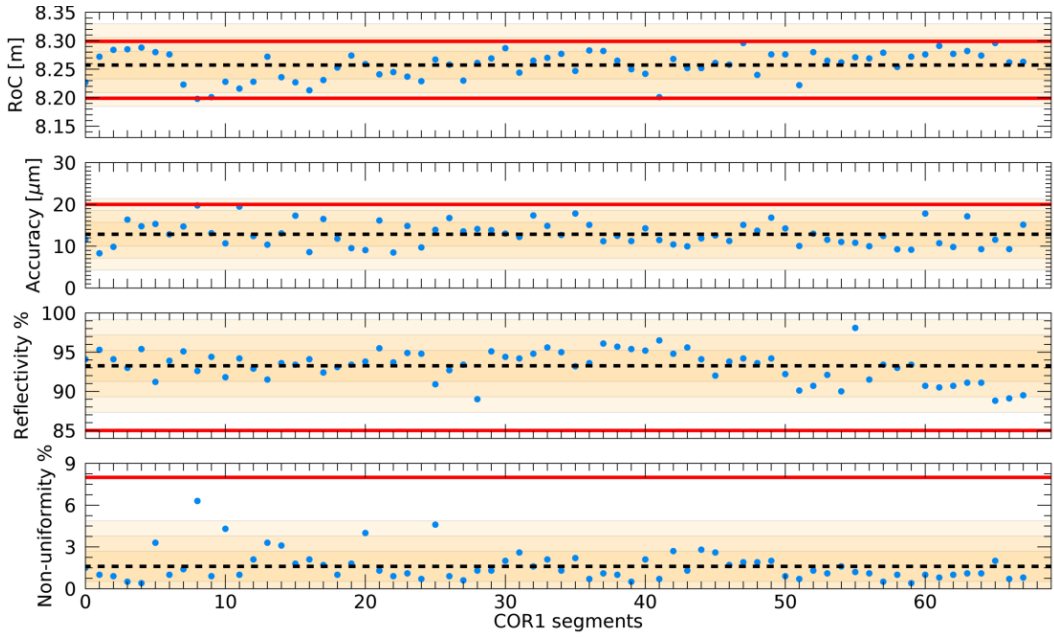

**Figure 10.** Main parameters of the COR1 segments of ASTRI M1, from the first (segment 0) to the last made (segment 67): radius of the spherical component (*upper panel*), residual shape error (*second panel*), average reflectivity (*third panel*), and reflectivity non-uniformity (*bottom panel*). For each panel the solid red and dashed black lines represent, respectively, the applicable requirement value and the average value.

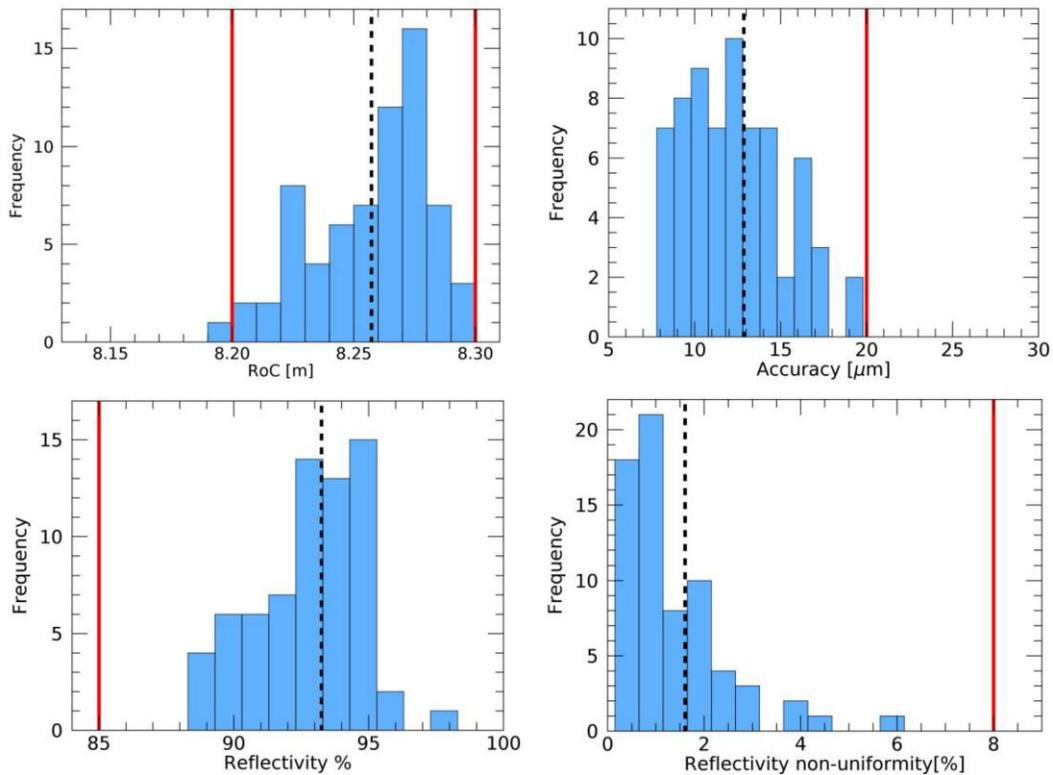

**Figure 11.** Distribution of the main parameters for the COR1 segments: radius of curvature (*upper left*), residual shape error (*upper right*), average reflectivity (*lower left*), and reflectivity non-uniformity (*lower right*). For each panel the red solid and black dashed lines represent, respectively, the applicable requirement value and the average value.

In Fig. 12 we show one example of the five reflectivity curves measured for each segment at five equidistant positions along a diagonal. Although these curves differ from each other, all of them are always above the required value of 85% between 300 and 550 nm, while they decrease at shorter wavelengths, where the data are reported only for information. They have a maximum at about 360 nm, while they show a gradual decrease at longer wavelengths. It should be noted that, in the case of ASTRI, the concept of focused reflectivity is not applicable, since we are talking of a dual-mirror system with the segments of the primary not having a real focus. Finally, in Fig. 13 we report the residual map obtained by comparing the theoretical and the real shapes of one ASTRI M1 segment. It is used to measure the surface accuracies reported in the second panel of Fig. 10.

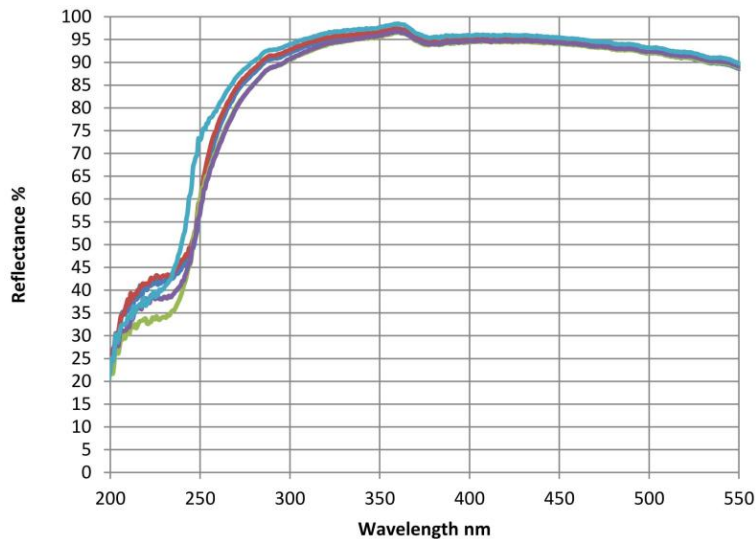

**Figure 12.** Example of the reflectivity curves of one ASTRI M1 segment measured at five different positions along a diagonal.

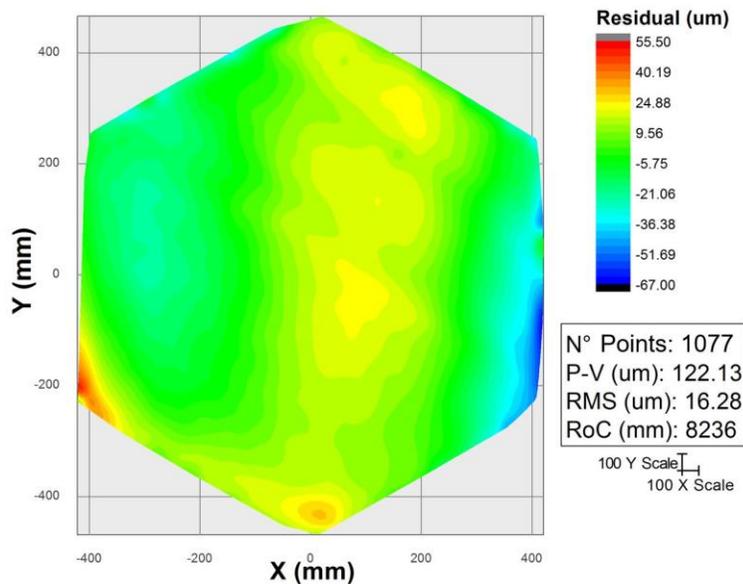

**Figure 13.** Example of the residual map of one ASTRI M1 segment.

## 9 Summary and Conclusions

Before starting the mass production, the manufacturing process of the mirror segments, based on the cold replication process set-up by INAF and ML, has been fully qualified for MST and ASTRI M1 mirrors, in the context of the CTA and ASTRI Mini-Array development phases. The ASTRI Mini-Array telescopes also represent the baseline design for the SST telescopes of CTA, which will be just slightly upgraded. It should be reminded that INAF will provide the entire set of mirrors of the 9 MST telescopes and the 37 SST telescopes for the CTAO northern and southern sites respectively, with about 1600 segments to be produced.

In the present work it has been verified that the developed manufacturing process ensures the realization of mirror segments that are fully compliant with the CTA requirements from the environmental, reliability, and performance points of view. We have also implemented an effective Quality Assurance approach that allows us to monitor the stability and reliability of the manufacturing process along the mass production.

The mirror production was completed at the end of 2019, with the delivery of 200 MST segments and 200 ASTRI M1 segments fully compliant with the requirements. Since a set of 10 ASTRI M2 mirrors is already available, not only more than enough mirrors are ready for the upcoming realization of the ASTRI Mini-Array, but a mass production exercise has been successfully carried out in view of the production of the mirrors for the SST CTAO array. In this respect, as the mirrors were highly aspherical, the optical performance were verified by means of metrological measurements, while the reflectivity performance was successfully verified with local measurements on different positions on the surface of each segment.

It should be noted that, in order to evaluate the optical performance of the MST mirrors, the qualification was carried out at the production premises of ML using both metrological and optical measurements. Moreover, other optical verifications were performed on a limited number of mirrors using other external facilities operated by INAF, DESY, and IAAT. All the results achieved are in good agreement. The reflectivity was measured on local areas of each mirror and, for a limited number of pieces, also in terms of focused reflectivity. Once again, the results are in good accordance. This means that the production process is stable and ensures the production of mirrors conforming to specifications. Taking into account the systematic effects of the different tests, the measurements carried out independently on sample mirrors confirmed the results obtained at the production site. Therefore, the results obtained with the described pilot production confirm that the acceptance criteria adopted by the industrial supplier are reliable. Based on this result, the current baseline for the future productions is to maintain the standard set of acceptance tests carried out at ML (reported in the flow chart of Fig. 4), with the only exception of the metrological measurements performed with the CMM. Since this type of measurements are particularly time consuming, they will be performed only on sample mirrors, in order to monitor the production process. On the other hand, we will still perform independent verifications at OAB and DESY laboratories, but on a reduced percentage of mirrors compared with the ~ 10 % adopted for the pilot production.

*Acknowledgments*

Part of the results reported in this paper were anticipated in 2019 at the SPIE Conference 11119 "Optics for EUV, X-Ray, and Gamma-Ray Astronomy IX", and were reported in the relevant SPIE Proceedings (La Palombara et al. 2019, SPIE Proc. 111191U). This work has been supported by the Italian Ministry of University and Research (MUR) with funds specifically assigned to the Italian National Institute of Astrophysics (INAF) for the Cherenkov Telescope Array (CTA), and by the Italian Ministry of Economic Development (MISE) within the "Astronomia Industriale" program. We acknowledge support from the Brazilian Funding Agency FAPESP (Grant 2013/10559-5) and from the South African Department of Science and Technology through Funding Agreement 0227/2014 for the South African Gamma-Ray Astronomy Programme. This work was conducted in the context of the CTA SST and MST Working Groups. We gratefully acknowledge financial support from the agencies and organizations listed here: http://www.cta-observatory.org/consortium_acknowledgments. We thank Dr. G. Tagliaferri, who coordinates the INAF participation to the MST Consortium, for his support to this activity. We thank Dr. G. Grisoni (ML) for providing us with useful comments and suggestions. This work has gone through internal review by the ASTRI Collaboration and by the CTA Consortium.

**Appendix: Results for ASTRI COR2 and COR3 segments**

In Fig. 14 and 15 we report the results obtained for, respectively, the ASTRI COR2 and COR3 segments.

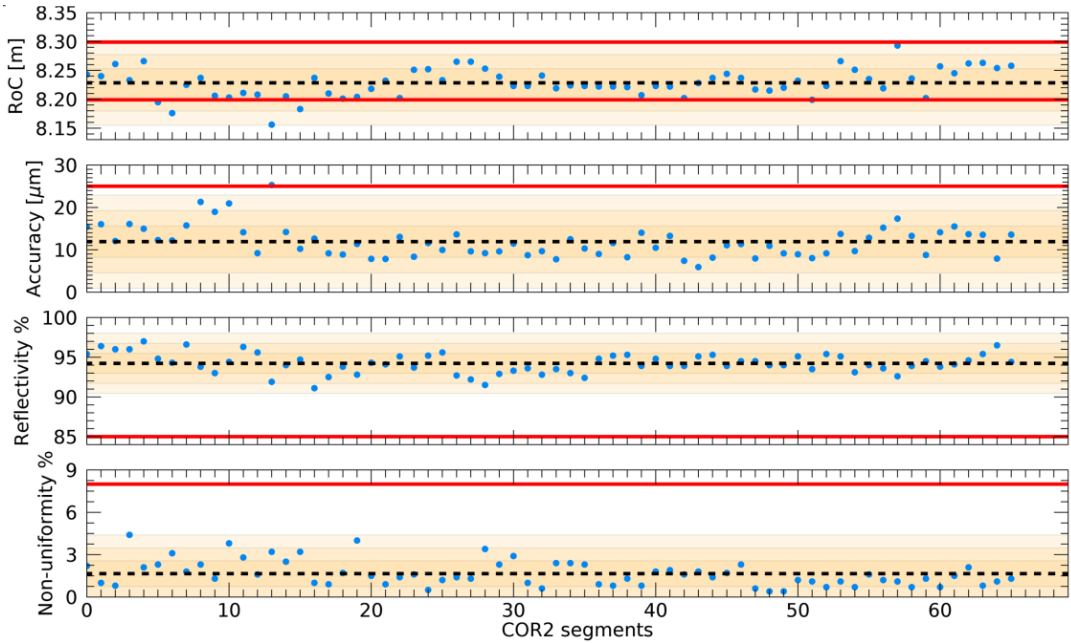

**Figure 14.** Main parameters of the COR2 segments of ASTRI M1, from the first (segment 0) to the last made (segment 65): radius of the spherical component (*upper panel*), residual shape error (*second panel*), average reflectivity (*third panel*), and reflectivity non-uniformity (*bottom panel*). For each panel the solid red and dashed black lines represent, respectively, the applicable requirement value and the average value.

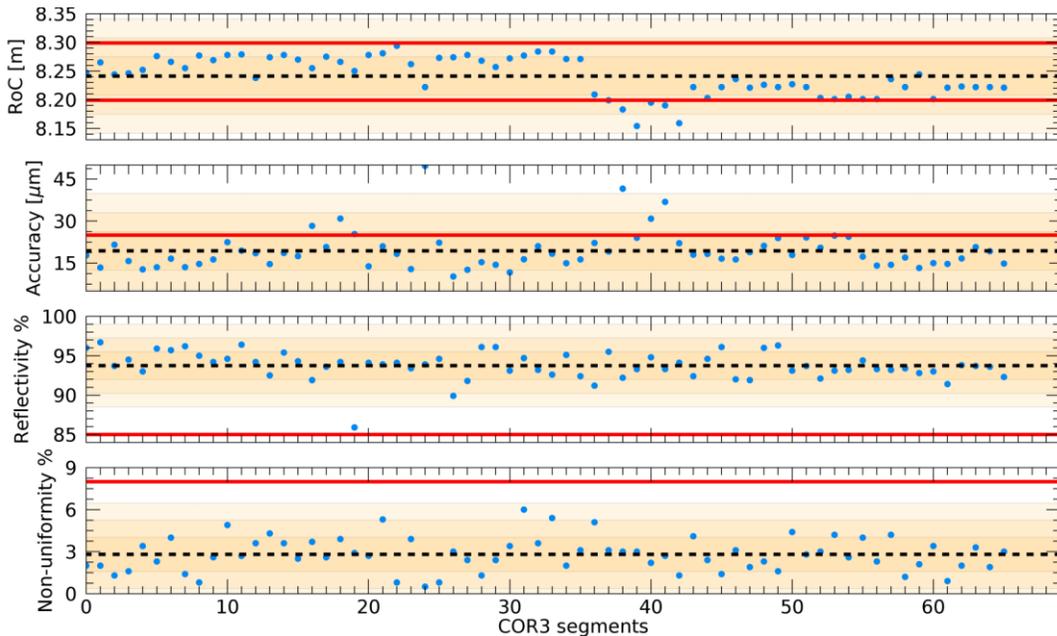

**Figure 15.** Main parameters of the COR3 segments of ASTRI M1, from the first (segment 0) to the last made (segment 65): radius of the spherical component (*upper panel*), residual shape error (*second panel*), average reflectivity (*third panel*), and reflectivity non-uniformity (*bottom panel*). For each panel the solid red and dashed black lines represent, respectively, the applicable requirement value and the average value.

In the first and second panel of each figure we show, respectively, the values of the RoC and of the residual shape errors. The first parameter is within the required range (between 8198 and 8298 mm) for nearly all the segments. The only exceptions are 4 COR2 segments and 5 COR3 segments, for which RoC < 8198 mm. Regarding the COR3 segments, we can note that the first 35 segments have a RoC systematically larger than the following ones. This change is due to the replication mould, which was reworked after the production of the first 35 segments. The second parameter is > 25 μm RMS for 1 and 7 segments, respectively. Moreover, regarding the COR2 segments, we can note that the residual shape error of the first 13 segments is larger than for the other segments, where the accuracy improved thanks to a better tuning of the replica process. For the segments which are not compliant with the requirement on the RoC and/or the residual shape error, it was necessary to perform a ray-tracing analysis of the shape data in order to assess that the segment shape is correct. In fact, even mirrors characterized by residual shape errors larger than the requirement can fulfil the telescope optical performance requirement. This is possible because the telescope optical performance requirement is defined on the angular dimension of the PSF across the field of view, while the requirement on the residual shape error is an extrapolation of such definition to mirror shape error heights. The measured shape of the panels out of specification was used as input to simulate with a ray-tracing code the PSFs of the panels at the focal plane, assuming the ASTRI design for the secondary mirror and detector. In this way we evaluated the optical performance degradation due to the primary mirrors, and verified that also these panels can be accepted.

Regarding the mirror reflectivity, in the third and fourth panel we report, respectively, its average value and its non-uniformity for each segment. As for the MST segments, also for all the ASTRI COR2 and COR3 segments these parameters are compliant with the lower limit of 85 % and the upper limit of 8 %, respectively.

In Fig. 16 and 17 the previous results are summarized in histograms, which report the distribution of each parameter around its average value.

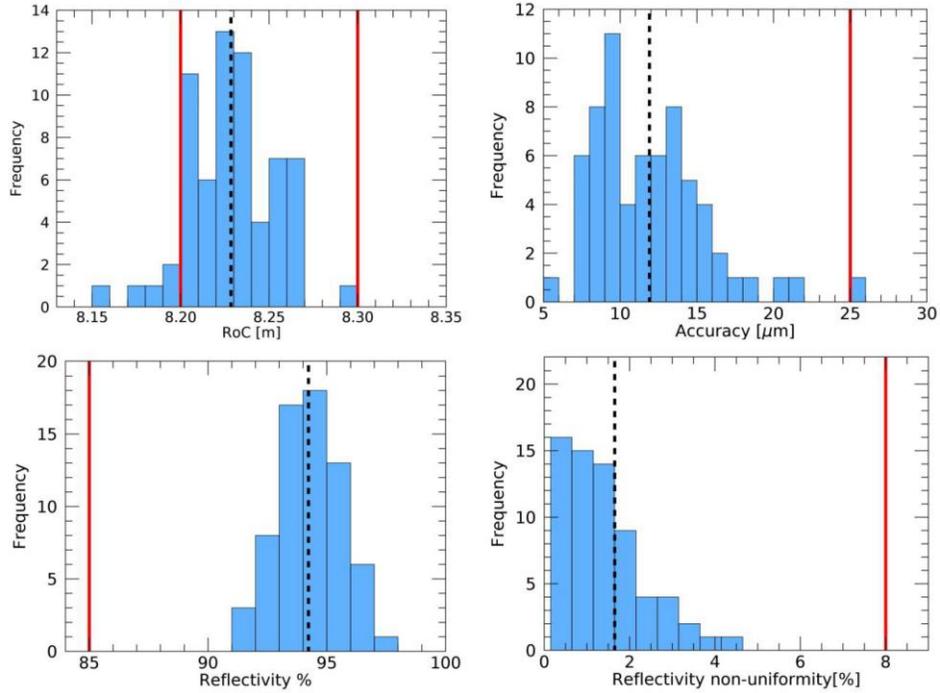

**Figure 16.** Distribution of the main parameters for the COR2 segments: radius of curvature (*upper left*), residual shape error (*upper right*), average reflectivity (*lower left*), and reflectivity non-uniformity (*lower right*). For each panel the red solid and black dashed lines represent, respectively, the applicable requirement value and the average value.

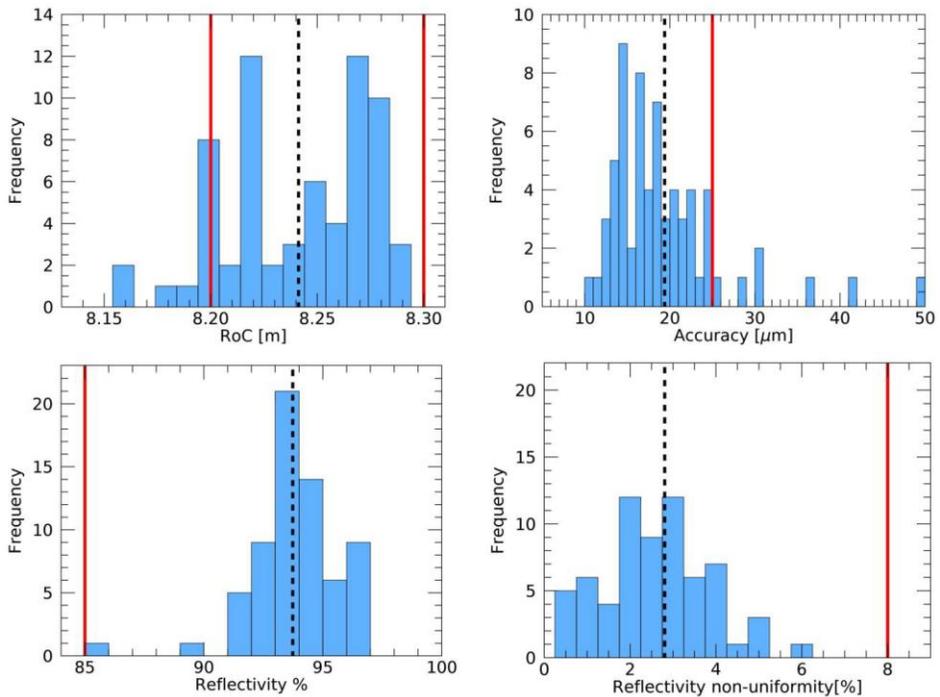

**Figure 17.** Distribution of the main parameters for the COR3 segments: radius of curvature (*upper left*), residual shape error (*upper right*), average reflectivity (*lower left*), and reflectivity non-uniformity (*lower right*). For each panel the red solid and black dashed lines represent, respectively, the applicable requirement value and the average value.